\documentclass{aa}

\usepackage{graphicx} 
\usepackage{txfonts} 
\usepackage{natbib} 
 
\newcommand{\Mpc}{$h^{-1}$\thinspace Mpc}

\def\apj{ApJ} 
\def\apjl{ApJL} 
\def\apjs{ApJS} 
\def\aj{AJ} 
\def\aap{A\&A} 
\def\mnras{MNRAS}

\begin{document}    
 
\title{The cosmic web for density perturbations of various scales } 
 
\author{ I. Suhhonenko\inst{1} \and J. Einasto\inst{1,2,3} \and 
  L. J. Liivam\"agi\inst{1} \and E. Saar\inst{1,2} \and 
  M. Einasto\inst{1}  \and  G. H\"utsi\inst{1} \and V. M\"uller\inst{4} \and 
  A. A. Starobinsky\inst{5,6} \and E. Tago\inst{1} \and  E. Tempel\inst{1} } 
 
\institute{Tartu Observatory, EE-61602 T\~oravere, Estonia 
\and 
Estonian Academy of Sciences,  EE-10130 Tallinn, Estonia
\and  
ICRANet, Piazza della Repubblica 10, 65122 Pescara, Italy 
\and 
Astrophysical Institute Potsdam, An der Sternwarte 16, D-14482 Potsdam, Germany 
\and 
Landau Institute for Theoretical Physics, Moscow 119334, Russia 
\and 
Research Center for the Early Universe (RESCEU), Graduate School of Science, 
The University of Tokyo, Tokyo 113-0033, Japan 
} 
 
\date{ Received; accepted}  
 
\authorrunning{I. Suhhonenko et al.} 
 
\titlerunning{Evolution of the cosmic web} 
 
\offprints{I. Suhhonenko, e-mail: ivan@aai.ee} 
 
\abstract {} 
{We follow the evolution of galaxy systems in numerical
  simulations.  Our goal is to understand the role of density
  perturbations on various scales in the formation and evolution of the
  cosmic web.} 
{We perform numerical simulations with the full power
  spectrum of perturbations, and with a spectrum cut at long
  wavelengths. In addition, we have one model, where we cut the
  intermediate waves. We analyse the density field and study void
  sizes and density field clusters in different models.}
{Our analysis shows that the fine structure (groups and clusters of
  galaxies) are created by small-scale density perturbations of scale
  $\leq 8$~\Mpc. Filaments of galaxies and clusters are created by
  perturbations of intermediate scale from $\sim 8$ to $\sim 32$~\Mpc, and
  superclusters of galaxies by larger perturbations. }
{We conclude that the texture of the cosmic web is
  determined by density perturbations of the scales up to $\sim 100$~\Mpc.
  Larger perturbations do not change the texture of the web, but
  modulate the richness of galaxy systems, and make voids emptier.}

\keywords {large-scale structure of Universe -- Cosmology:
  miscellaneous -- Cosmology: theory -- Galaxies: clusters: general --
  dark matter -- Methods: numerical}

\maketitle

\section{Introduction} 
 
The first studies of the three-dimensional distribution of galaxies 
and clusters of galaxies demonstrated that they are not distributed 
randomly, but are mostly located in filamentary superclusters connected by galaxy 
filaments to form a connected network, leaving large regions devoid of 
galaxies \citep{Gregory:1978, Joeveer:1978pb, 
  Joeveer:1978dz,Tarenghi:1978, Tully:1978, Einasto:1980, 
  Zeldovich:1982kl}.  This observational picture was rather similar to 
the theoretical prediction of \citet{Zeldovich:1970}, the so-called pancake 
scenario of galaxy formation, as discussed by 
\citet{Zeldovich:1978}. However, there were some important differences 
between the models and the observations. In the real world, the most common 
structural elements were filaments of galaxies and their clusters, whereas the 
\citet{Zeldovich:1978} scenario predicted the formation of flat 
formless pancake-like systems (for a discussion see 
\citet{Zeldovich:1982kl} and \citet{Einasto:1983qa}).  The simulations 
performed for the neutrino-dominated universe, called the Hot 
Dark Matter (HDM) model \citep{White:1983}, had similar problems. 
 
The problem of the absence of the fine structure was solved when the 
Cold Dark Matter (CDM) scenario was suggested by \citet{Bond:1982} and 
others.  The basic difference between the HDM and CDM scenarios is the 
absence or presence of small-scale density fluctuations. The 
comparison of simulations for both dark matter types demonstrated 
the advantages of the CDM model -- in this model, the fine structure is present in 
the form of filaments of various scales (see 
\citet{Melott:1983} and \citet{White:1987} among others). An even 
closer agreement of simulations with observations is obtained in 
models with a cosmological constant, $\Lambda$CDM, as demonstrated by 
\citet{Gramann:1988}. 
 
One problem remained: both numerical simulations and direct 
observations suggested that the dominant structural elements of the 
cosmic web are filaments of various scales, whereas the classical 
\citet{Zeldovich:1970} scenario predicted the formation of flat 
pancake-like systems.  This problem was solved when 
\citet{Bond:1996fv} studied in detail the formation of the large-scale
structure. They found that primordial tidal fields play an important 
role in the evolution of structure, thus the dominant elements are 
filaments, not flat pancakes. 
 
\citet{Bond:1996fv} showed why the cosmic web has a filamentary
character, but could not explain, why there is a large variety of
systems from small groups and weak filaments to rich clusters and
superclusters. It is clear intuitively that the main reason for the
formation of systems of various richnesses is the presence of density
perturbations of different scales.  This idea is justified by
comparing the HDM and CDM scenarios -- the fine structure of the
cosmic web is generated by density fluctuations of medium and small
scales that are absent in the HDM model.
 
\begin{figure*}[ht] 
\centering 
\resizebox{0.8\textwidth}{!}{\includegraphics{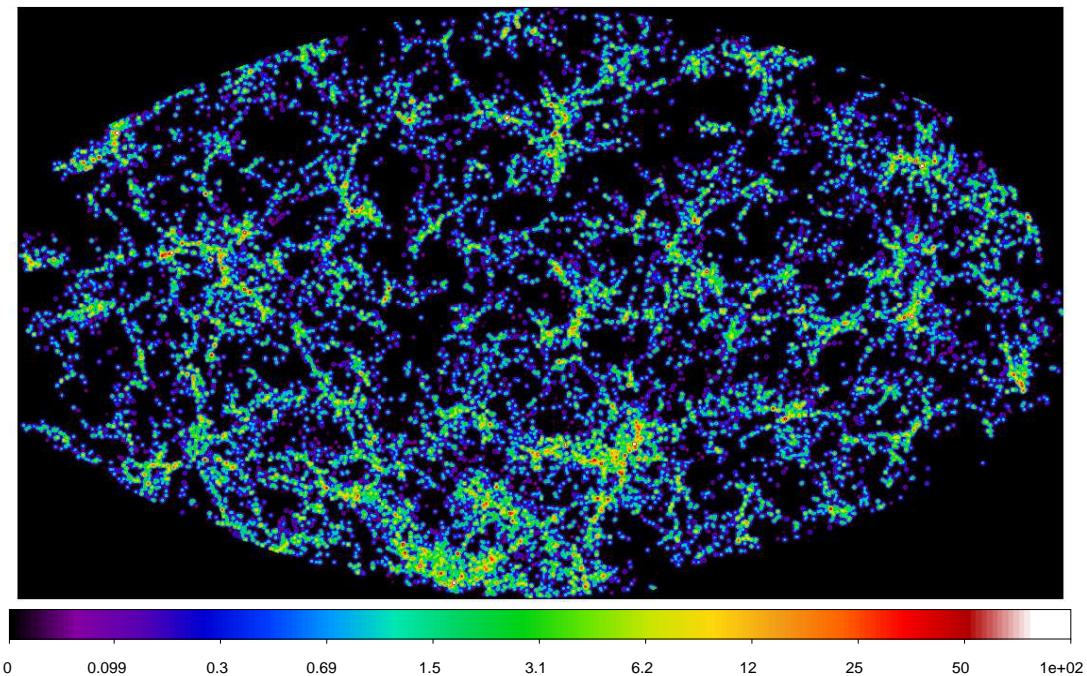}} 
\caption{The luminosity density field of the SDSS in a spherical shell
  of 10~\Mpc\ thickness at a distance of 240~\Mpc.  To enhance the faint
  filaments in voids between the superclusters, the density scale is
  logarithmic, in units of the mean luminosity density for
  the whole DR7.  The rich complex in the lower area of the picture is
  part of the Sloan Great Wall; it consists of two very rich
  superclusters, SCL~111 and SCL~126 in the list by
  \citet{Einasto:2001oq}.  The coordinates $x,y$ are defined by eqs. (1)
  and (2).  A cellular distribution of superclusters and filaments of
  various richness can be clearly seen. }
\label{fig:dr7_240} 
\end{figure*}

The role of very large density waves on the formation of the structure
has been studied by a number of authors.
\citet{Knebe:2000eu}, \citet{Bagla:2006gf}, and \citet{Bagla:2009dq}
investigated the effect of the resolution of cosmological simulations on
the properties of the filamentary web.  \citet{Power:2006ai} studied the
impact of the simulation box size on the properties of the simulated
cosmic web. \citet{Bernardeau:2002ff} gave a review of how 
non-linear perturbation theory helps us to understand the large-scale
structure of the Universe and various statistical tools used to
characterise its properties.
 
It is well known that quasi-relaxed systems of galaxies, such as
groups and clusters, are formed by perturbations, which have entered
the non-linear stage of their evolution.  The scale of the 
non-linearity is usually taken equal to about 8~\Mpc.  The
richest non-percolating systems of galaxies are superclusters. Rich
superclusters form a cellular distribution, with large voids 
surrounded by rich superclusters.  The characteristic diameter of
these supervoids is of the order of 100~\Mpc\
\citep{Einasto:1994wd,Einasto:1997lh}.  

 Supervoids are not empty but contain a hierarchy of
  voids, as demonstrated by many authors
  \citep{Martel:1990,Lindner:1995ui,Gottlober:2003,
    Aragon-Calvo:2007,von-Benda-Beckmann:2008qf,
    van-de-Weygaert:2009pd,van-de-Weygaert:2009qf, Aragon-Calvo:2010,
    Jones:2010cr}.  It is natural to expect that the web of rich
  superclusters is formed by density perturbations of scale 100~\Mpc\
  and above.  The skeleton of the cosmic web is discussed by,
 among others, \citet{Sheth:2004ly}, \citet{Hahn:2007ve},
  \citet{Forero-Romero:2009}, \citet{Sousbie:2008ul, Sousbie:2009lq},
  \citet{Aragon-Calvo:2010wd}, \citet{Bond:2010cr,Bond:2010rr}, and
  \citet{Einasto:2010b}. Thus, we can say that the role of
small-scale and large-scale density perturbations is relatively well
understood.  However, it remains unclear how perturbations of
intermediate scale combine to form elements of the cosmic web.
 
We aim to investigate in greater detail than before the role of
density perturbations of different scales in the formation of the
cosmic web. We perform numerical simulations of the formation and
evolution of the large-scale structure, using different box sizes and
resolutions. Furthermore, to elucidate the influence of perturbations
of different scales, a varying large-scale cutoff is introduced into
the power spectrum of initial perturbations. All simulations of a
given series have identical initial conditions with random initial
positions and velocities of test particles, but the amplitude of all
perturbations on a scale exceeding a given one is forced to be
zero. In this way, we can follow how systems of galaxies grow under
the influence of perturbations of various scales.
 
To characterise the effect of varying the cutoff scale of density 
perturbations, we consider three quantitative tests: the mean radii 
of voids defined by groups and clusters of galaxies,  the mass functions of 
clusters of galaxies, and the density distributions of particles in 
the void and the supercluster core regions. 
 
The paper is organised as follows. In the next section, we describe  
numerical simulations used to model the evolution of the cosmic 
web.  For comparison, we also calculate the SDSS luminosity density 
field.  In Section 3, we analyse the cluster mass functions and void sizes. 
We analyse the density distributions in the void and the supercluster core 
regions and discuss our results in Section 4.  The last section presents 
the summary of the analysis.

\section{Modelling the  evolution of the cosmic web} 
 
\subsection{The SDSS luminosity density field} 
 
To compare simulated void and cluster data with actual data, we used 
the recently completed Sloan Digital Sky Survey (SDSS) Data Release 7 
(DR7)  (see \citet{Abazajian:2009th}).  In the present analysis, we used 
only the contiguous northern zone of the DR7. The analysis was made in 
several steps. First we calculated the luminosity density field. We 
estimated the total luminosities of groups and isolated galaxies in a 
flux-limited sample, using luminosity  weights that take into 
account galaxies and galaxy groups too faint to fall into the 
observational window of absolute magnitudes at the distance of the 
galaxy.  For details of the data reduction, we refer to \citet{Tago:2010ij}
and \citet{Tempel:2011}.  The high-resolution luminosity density field 
was calculated with the $B_3$ spline of kernel size 1~\Mpc. 
 
The high-resolution luminosity density field was found using spherical 
coordinates (the SDSS coordinates $\eta$ and $\lambda$) and distance.  To 
obtain an impression of the luminosity density field of the whole 
northern region we show in Fig.~\ref{fig:dr7_240} the luminosity 
density field in a spherical shell at a distance of 240~\Mpc\ from ourselves. It 
is based on the SDSS angular coordinates $\eta$ and $\lambda$, and 
the comoving distances $d_{\mathrm{gal}}$ of galaxies 
\begin{eqnarray} 
x&=&-d_{\mathrm{gal}} \lambda,\\ 
y&=&d_{\mathrm{gal}}  \eta \cos\lambda,\\ 
z&=&d_{\mathrm{gal}} - d_0, 
\end{eqnarray} 
where $d_0=50$~\Mpc\ is the minimal distance used in the calculation of
the density field. A supercluster catalogue based on the luminosity
density field of the full contiguous northern SDSS region was
published by \citet{Liivamagi:2010}.  The representation in spherical
coordinates serves basically for illustrative purpose, as it is
practically free of distance dependent selection effects.  The
spherical shell has a thickness of 10~\Mpc\ to enhance the filamentary
distribution of galaxies, clusters, and superclusters.  The mean
distance of the shell corresponds to the distance of the Sloan Great
Wall, seen in the lower area of the figure. We see that it consists 
of two very rich superclusters: SCL111 and SCL126 according
to the catalogue by \citet{Einasto:2001oq}.

\subsection{Simulation of the cosmic web for density perturbations of 
  various scales}  
 
To understand the formation of the filamentary
supercluster-void web correctly, we must perform numerical
simulations for a box that contains both small and large
waves.  The smallest units of this network are galaxies, the most
frequent systems of galaxies are groups and clusters of galaxies.  The
characteristic scale of groups is 1~\Mpc\ (galaxies are still about 10
to 100 times smaller), thus the simulation must have a resolution of at least 
this scale.  On the other hand, the largest
non-percolating systems of galaxies are superclusters, which have a
characteristic scale of 100~\Mpc\ \citep{Oort:1983, Zucca:1993,
  Einasto:1994wd,Kalinkov:1995,
  Einasto:1997nx,Einasto:2001oq,Erdogdu:2004, Einasto:2007tg}.
Superclusters have rather different richness, from small systems like
the Local Supercluster to very rich systems such as the Shapley
Supercluster.  It is clear that this variety has its origin in density
perturbations of still larger scales.  Thus, to understand
the supercluster-void phenomenon correctly, the influence of very
large density perturbations should also be studied.
 
To derive both a high spatial resolution and the presence of density
perturbations on a large scale interval, we used a number of
simulations in boxes of sizes from 100~\Mpc\ to 768~\Mpc, and various
resolutions of  $N_{\mathrm{grid}}^3 = 256^3$ and $N_{\mathrm{grid}}^3
= 512^3$ particles and simulation cells.  To see the effect of
perturbations of various scales, we use simulations with the full
power spectrum, as well as with a power spectrum truncated at
wave-numbers $k_{\mathrm{t}}$, so that the amplitude of the power
spectrum on large scales is zero i.e., $P(k) = 0$, if $k< k_{\mathrm{t}}$,
wavelength $\lambda_{\mathrm{t}} = 2\pi/k_{\mathrm{t}}$.

{
\begin{table}[ht] 
\caption{Parameters of the models.} 
\begin{tabular}{lrrrr}  
\hline  
\noalign{\smallskip}
Model   & $L$ & $\lambda_{\mathrm{cut}}$    & $N_{\mathrm{cl}}$ A  &$N_{\mathrm{cl}}$ B \\  
\noalign{\smallskip}
\hline  
&(1)&(2)&(3)&(4) \\ 
\noalign{\smallskip}
\hline  
\noalign{\smallskip}
\\ 
M768.768    & 768 & 768   &   102783  & \\ 
M768.128    & 768 & 128   &   105496  & \\ 
M768.032    & 768 &   32   &   184193  & \\ 
M768.012    & 768 &   12   &   213569  &\\ 
\\ 
M256.256    & 256 & 256 &     54224 & 5578  \\ 
M256.064    & 256 & 64  &     56000&  6077 \\ 
M256.032    & 256 & 32  &     60985&  6996\\ 
M256.016    & 256 & 16  &    70813 &   7624\\ 
M256.008    & 256 &  8   &     121022 &  125\\ 
M256.864    & 256 & 8-64 &          &        \\ 
\\ 
L256.256    & 256 &  256  &   341758   &  59100 \\ 
L256.128    & 256 &  128  &   343706   &  59313\\ 
L256.064    & 256 &   64  &    351198  &  61831  \\ 
L256.032    & 256 &   32  &    372298  &  68221 \\ 
L256.016    & 256 &   16  &    426699  &  84691 \\ 
L256.008    & 256 &    8   &    546414  &  126832 \\ 
\\ 
L100.100    & 100 &  100  &    54348  &  173655  \\ 
L100.032    & 100 &   32  &    57878  &  182675 \\ 
L100.016    & 100 &   16  &    62396  &  197053 \\ 
L100.008    & 100 &    8   &    71726  &  237499 \\ 
\noalign{\smallskip}
\hline
\label{Tab1}                         
\end{tabular} 
\tablefoot{
\\
\noindent Column 1: $L$ -- the size of the simulation box in \Mpc;\\ 
\noindent Column 2: $\lambda_{\mathrm{cut}}$ -- the cut-off scale in \Mpc;\\ 
\noindent Column 3: The number of DF clusters in the high-resolution 
density field at a redshift $z=0$, for the model L256 using the
parameter set A (see Sect. 2.5 for explanation);\\   
\noindent Column 4: The number of DF clusters in the high-resolution 
density field at a redshift $z=0$, for the model L256 using the parameter
set B;  the number of AHF halos at the redshift $z=0$ for the 
models M256 and  L100. }
\end{table} 
} 
 
All models of the same series have the same realisation of random
Fourier amplitudes, so the role of different waves in the models can
be easily compared.  To compare different models of a series, every
particle has an identification number, the same for all models of the
same series.  The main model parameters are given in
Table~\ref{Tab1}. In addition to models with cuts on large scales, we
calculated a model, where the density perturbations on intermediate
scales were truncated to zero at the wavelengths $8 - 64$~\Mpc, the
model M256.864. This model shows the effect of the absence of
perturbations of medium scales.
 
We use the following notation for our models: the first 
characters M and L designate models with resolutions of
$N_{\mathrm{grid}} = 256$ and $N_{\mathrm{grid}} = 512$, respectively; 
the following number gives the size of the simulation box, $L$, in 
\Mpc; the subsequent number indicates the maximum wavelength used in the 
simulation, also in \Mpc; if the full power spectrum is used these two numbers 
coincide.  The locations of the cells inside the cubical density grid 
are marked by cell indices $(i,j,k)$, where $i$, $j$, and $k$ are 
integers that run from $1$ to $N_{\mathrm{grid}}$.

 \begin{figure*}
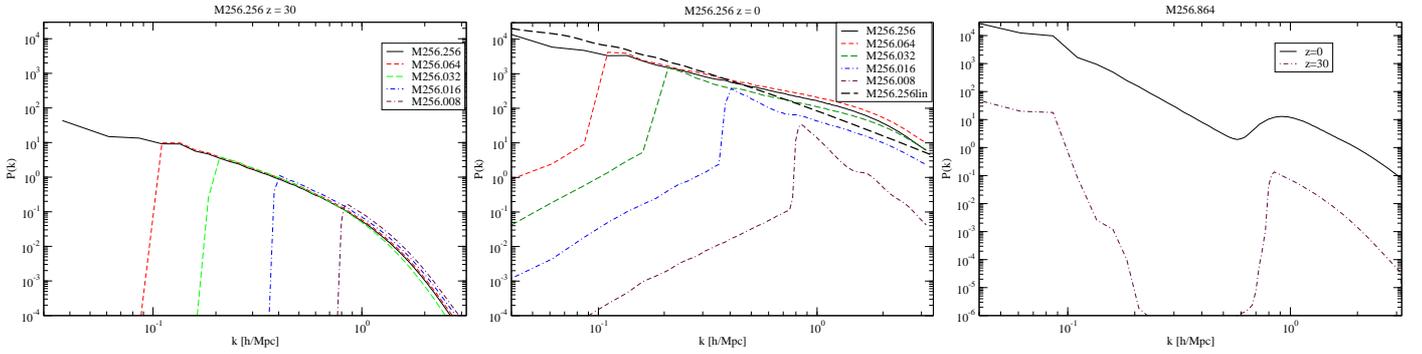
 
\centering 
\resizebox{0.33\textwidth}{!}{\includegraphics*{16394fg2a.eps}} 
\resizebox{0.33\textwidth}{!}{\includegraphics*{16394fg2b.eps}} 
\resizebox{0.33\textwidth}{!}{\includegraphics*{16394fg2c.eps}} 
\caption{The left and middle panels show the power spectra for the models of the
  series M256 at the epochs $z = 30$ and $z = 0.00$, respectively.  The
  right panel shows the power spectra for the model M256.864 at the epochs $z =
  30$ and $z = 0.00$.  In the middle  panel, we also show the
  power spectrum for the full model M256.256 at the present epoch $z=0$ where
 the evolution has been linear on all scales.}
\label{fig:spec} 
\end{figure*}

For the models of the M256 and the M768 series, we used in simulations
the AMIGA code, which is the follow-up of the multi-level adaptive particle
mesh (MLAPM) code by \citet{Knebe:2001qe}.  This code uses an adaptive
mesh technique in the regions where the density exceeds a fixed
threshold.  In this code, gravity is automatically softened adaptively,
so that the softening length is close to its optimum value in both high
and low-density regions.  We chose a maximum level of eight
refinements. For the models of the L100 and L256 series, we used the
GADGET-2 code with gravitational softening length
10~$h^{-1}$\thinspace kpc and 20~$h^{-1}$\thinspace kps, respectively 
\citep{Springel:2001,Springel:2005}. The simulations M256 and M768
were performed at the Tartu Observatory, the simulation L100 at the
Astrophysical Institute Potsdam, and the simulation L256 at the High
Performance Computing Centre of the University of Tartu.
 
The initial density fluctuation spectra were generated using the 
COSMICS code by \citet{Bertschinger:1995}\footnote{\tt http://arcturus.mit.edu/cosmics}. 
We assumed the cosmological parameters $\Omega_{\mathrm{m}} = 0.28$, 
$\Omega_{\Lambda} = 0.72$, $\sigma_8 = 0.84$, and the dimensionless 
Hubble constant $h = 0.73$; to generate the initial data, we used 
the baryonic matter density $\Omega_{\mathrm{b}}= 0.044$ 
(\citet{Tegmark:2004}). We generated initial conditions for a truncated
power spectrum, using the full spectrum to calculate the model parameters.
The amplitude of a spectrum was set to zero for $k< k_{t}$
during the calculation of the initial density field, keeping all
simulation parameters fixed across the full set of realisations.
The calculations were started for an early epoch, $z=100$.  The particle positions 
and velocities were extracted for 12 epochs between the redshifts 
$z = 30 \dots 0$ (the models of the M256 and the M768 series), and for 
7 epochs between the redshifts $z=100 \dots 0$ (the models of the L100 and
the L256 series).

\subsection{Calculation of the density field} 
 
For every particle, we calculated the local density in units of the 
mean density, using the positions of 27 nearby particles.  We also calculated 
the mean global density at the location of the particle.  For 
this purpose, we first determined the density field, using a $B_3$ spline 
\citep[see][]{Martinez:2002fu}: 
\begin{equation} 
B_3(x)=\frac1{12}\left[|x-2|^3-4|x-1|^3+6|x|^3-4|x+1|^3+|x+2|^3\right], 
\end{equation} 
where function differs from zero only in the interval $x\in[-2,2]$. 
The one-dimensional $B_3$ box spline kernel of width $h=N$ is 
\begin{equation} 
K_B^{(1)}(x;N)=B_3(x/N)/N. 
\end{equation} 
This kernel preserves the interpolation property (mass conservation) 
for all kernel widths that are integer multiples of the grid step, $h=N$. 
The 3D $K_{B}^{(3)}$ box spline kernel  is given by the direct  
product of three one-dimensional kernels 
\begin{equation} 
  K_B(\mathbf{x};N)\equiv K_B^{(3)}(\mathbf{x};N)=K_B^{(1)})(x;N)K_B^{(1)}(y;N) 
        K_B^{(1)}(z;N), 
\end{equation}  
where $\mathbf{x}\equiv\{x,y,z\}$.  
 
To calculate the high-resolution density field, we use the kernel of
the scale equal to the cell size of the simulation,
$h=L/N_{\mathrm{grid}}$, where $L$ is the size of the simulation box,
and $N_{\mathrm{grid}}$ is the number of grid elements in one
coordinate.  We use consecutive smoothing stages; the smoothing stage
of the order $i$ has the smoothing scale $r_i= L/N_{\mathrm{grid}}
\times 2^i$. The effective scale of smoothing is equal to $2\times
r_i$.  We applied this smoothing up to index (order) 6.  The smoothing
with the $B_3$ spline kernel is rather close to the smoothing with an
Epanechnikov kernel of the same scale, used in earlier studies of
superclusters \citep{Einasto:2006kl,Einasto:2007tg}. For the model
L100, the smoothing index 5 corresponds to the kernel of the radius
6.25~\Mpc, for the model L256 the smoothing index 4 corresponds to the
kernel of the radius 8~\Mpc, and for the model M768 the smoothing
index 2 corresponds to the kernel of the radius 12~\Mpc.  These kernel
radii were used to calculate the global density field used in the
selection of the void and supercluster core regions.
 
\begin{figure*}[ht] 
\centering 
\resizebox{0.80\textwidth}{!}{\includegraphics*{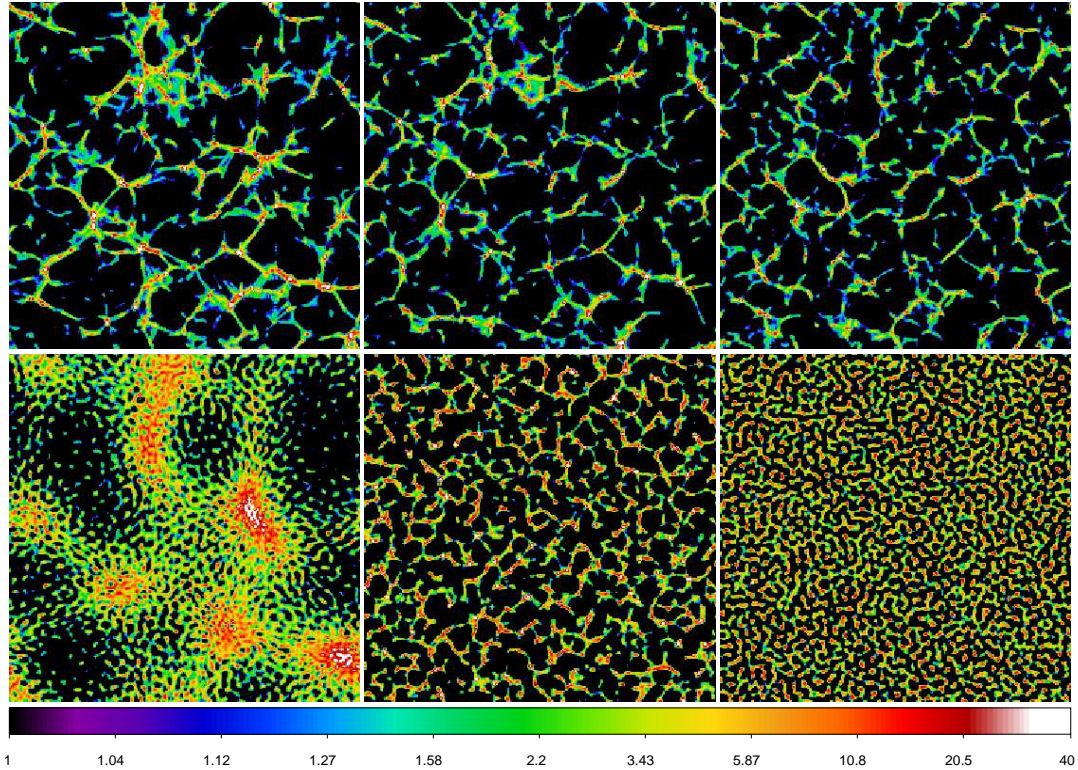}} 
\caption{Density fields for the models of the M256 series.  The upper 
  panels show the high-resolution fields for the models M256.256, M256.064, and 
  M256.032, the lower panels -- for the models M256.864, M256.016, and 
  M256.008 (from left to right).  The densities are shown for a layer of 
  6~\Mpc\ thickness at the $k=75$ coordinate. All fields correspond to the 
  present epoch $z=0$. The densities are expressed on a logarithmic scale to 
  enhance the low-density regions. Only the overdensity regions are shown. 
  The upper limits to the density are 40 in the 
  upper panels, and 10, 10, and 5 in the lower panels.} 
\label{fig:Mdenfield} 
\end{figure*}

The power spectra of the models of the series M256 are shown in
Fig.~\ref{fig:spec} for both an early epoch, $z = 30$, and the present
epoch $z = 0.0$. The power spectra for the model M256.864 are shown in
the right panel of the same figure for the same epochs. For
comparison, we also show in the middle panel the power spectrum of the
model M256.256 for the present epoch, if evolved linearly at all
scales. Fig.~\ref{fig:spec} shows that at an early epoch and at small
wavelengths all spectra practically coincide.  At the present epoch,
the amplitudes of spectra for different models differ at small
wavelengths: the models with the truncated spectra evolve at small
wavelengths more slowly than the full model.  At small wavelengths,
the slowest evolution is seen in the model M256.008, where the growth
of the amplitudes is almost linear on all scales.  This effect is
expected because density waves of larger size that could amplify the
growth of small-scale waves are absent (see the analysis below).  We
also see that in all models with a cutoff, the amplitude of the
spectrum in the cutoff range slowly increases, i.e.\ some power is
translated from smaller to larger waves. This effect is much stronger
in the model M256.864, where in the cutoff range of the spectrum a
fairly high amplitude of the spectrum develops for the present epoch,
only a factor of ten lower than the amplitude at the cutoff wavelength
$\lambda = 8$~\Mpc.

The high-resolution density fields at the present epoch for models of
the M256 series are shown in Fig.~\ref{fig:Mdenfield}.

\subsection{Comparison of halo finders}

In simulations, instead of groups and clusters of galaxies, 
  dark matter halos are usually used to define simulated compact systems.  In
  the pilot phase of this study, we used the PM code in a box of size
  $L=128$~\Mpc\ with $128^3$ particles and simulation cells.  For
  these models, we applied the conventional halo finder using the FoF
  technique with the linking length parameter 0.2 in units of the mean
  separation of particles. The analysis showed that the FoF halos are of
  little use for our study, since in the central regions of superclusters,
  the richest FoF halos contain the whole core of the supercluster, and have
  sizes of the order of 10~\Mpc.

  For this reason, we tried the adaptive Amiga Halo Finder (AHF) code
  developed by \citet{Knollmann:2009}, where the number of particles
  in a halo is $N_{\mathrm{p}} \geq 20$.  Halos and their parameters
  (masses, virial radii, positions, velocities etc.)  were found for
  most models and simulation epochs.  Here we have another difficulty:
  the number of halos for early epochs (high redshifts) decreases
  dramatically with increasing $z$: there are about 1000 times fewer
  halos at the redshift $z=10$ in the model L100 than at the present
  epoch.  For models of the M256 series, the AHF finds no halos at the
  redshifts $z=5$, and $10$, about 10 times fewer halos than the
  density field cluster method (see below) for the models M256.256,
  M256.064, M256.032, and M256.016, and almost no halos for the model
  M256.008 (see Table~\ref{Tab1}).  The density enhancements in these
  models are much smaller and do not conform to the definition of the
  AHF halos, (which have at least 20 particles), in spite these models
  containing many small compact density enhancements at all
  redshifts. In Fig.~\ref{fig:Mdenfield}, we present the density
  fields of the models of the M256 series at the redshift $z=0$, and
  Figs. 1--3 of \citet{Einasto:2010b} for higher redshifts.
 
Taking these difficulties into account, we used the high-resolution 
density field to define compact systems, i.e., density field (DF)
clusters.  DF clusters correspond to groups and clusters of galaxies in the 
real Universe, and to halos in simulations.  In defining density field 
 clusters, we applied the conventional tradition, used by Abell in 
his rich cluster catalogues \citep{Abell:1958bs, Abell:1989fy}.  The Abell 
clusters were selected using galaxies in a sphere of a radius of about 
1.5~\Mpc. Our density field method to find halos is rather similar to
the DENMAX method by \citet{Bertschinger:1991}
and \citet{Gelb:1994}. The basic difference lies in the detail:  we do not use
particles to define the center and the mass of the halo, but only data
obtained directly from the density field.  The procedure for finding the DF
clusters is described in the next subsection.

\begin{figure*}[ht]
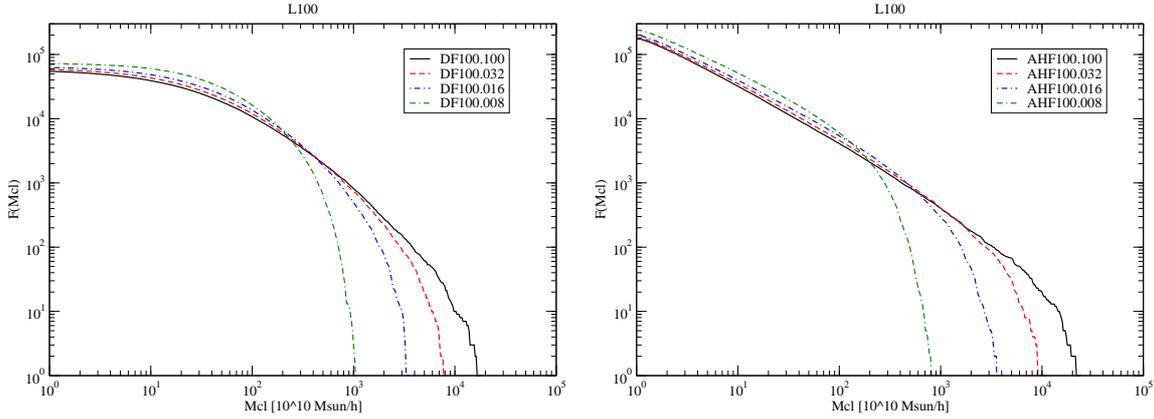
 
\centering 
\resizebox{0.40\textwidth}{!}{\includegraphics*{16394fg4a.eps}}
\hspace{2mm}  
\resizebox{0.40\textwidth}{!}{\includegraphics*{16394fg4b.eps}}
\hspace{2mm} 
\caption{The left panel shows the cumulative mass functions of the density 
  field clusters for the model L100 for various cutoff scales. The right panel
  shows  the cumulative mass functions of the AHF halos for the same model.
   } 
\label{fig:L100_massf} 
\end{figure*}

 To compare our DF cluster mass distributions with the mass
  distributions of the AHF halos, we applied both methods for the highest
  resolution model L100.  We found the DF clusters using the parameters
  $D_0=2$ (in units of the mean density), $D_{\mathrm{p}} =20$ (the
  minimal number of particles), and $N=8$; the latter
  corresponds to the virial (outer) radius of the halo
  $R_{\mathrm{vir}}=1.66$~\Mpc, which is similar to the virial radii of the
  largest halos found with the AHF method.  The cumulative mass functions
  for the DF clusters and the AHF halos for all models of the L100 series for
  the redshift $z=0$ are shown in Fig.~\ref{fig:L100_massf}. 

  Fig.~\ref{fig:L100_massf} shows that the maximum masses of halos
  for both halo finders are practically identical for all models with
  different cutoffs.  The shapes of the cumulative mass functions are
  slightly different. Table~\ref{Tab1} and Fig.~\ref{fig:L100_massf}
  show that the total numbers of halos found with the AHF method are
  about three times larger than the numbers of the DF clusters for models with
  the same cutoff. This difference is expected, since AHF treats all
  subhalos as independent halos.

  In the mean mass range, the number of the DF clusters is a bit higher
  than the number of the AHF halos.  This difference is also
  expected. Our DF cluster finder always has an identical volume around
  a high-density peak, whereas the virial radii of the AHF halos
  decrease for the model L100.100 from $R_{\mathrm{vir}} = 1.5$~\Mpc\
  to $R_{\mathrm{vir}} =0.05$~\Mpc\ for the considered mass range.  For
  this reason, close intermediate-mass halos are counted as separate
  small halos in the AHF method, but as a single larger halo in the DF
  cluster method.

These differences in the mass distributions are not important for the
present study, since we compare mass distributions for different cutoffs of
the power spectra, obtained with the same halo identification
method. We use halos basically to identify voids and to measure their
radii; the internal structures of the halos are not important for the
present analysis.

\subsection{Density field clusters} 
  
As a first step, we find local density maxima of the high-resolution
density field.  In the search for maxima, we use three parameters: the
minimum density threshold $D_0$, above which we search for density
maxima, and the minimum mass of the DF clusters, $D_{\mathrm{p}}$.  During
the search, densities are expressed in units of the mean density of the
simulation, and masses -- in units of the particle mass of the
simulation (the number of particles in a halo). The
mass of the DF clusters, $M_{\mathrm{cl}}$, is calculated by adding
the local densities in the cells within $\pm N$ cells from the central
one, i.e.\ in a total in $(2N+1)^3$ cells.  The number $N$ defines the
volume where we count the mass of the cluster, and it is the third parameter
of the search. We express the DF cluster masses in solar units, using
the known particle mass.

For the M256 model, we used the parameters $D_0=2$, $D_{\mathrm{p}}
=5$, and $N=3$; for the M768 model, we used the parameter set $D_0=2$,
$D_{\mathrm{p}} =5$, and $N=1$, which correspond to the cluster search
radii $r=3.5$~\Mpc\ and $r=4.5$~\Mpc, respectively.  To find the
dependence of the DF cluster mass functions and the cluster-defined
void radii on the parameters used in the definition of a DF cluster,
we used two sets of parameters for the models of the L256 series -- a
high-sensitivity set A with $D_0=2$, $D_{\mathrm{p}} =5$, and $N=3$,
and a low-sensitivity set B with $D_0=10$, $D_{\mathrm{p}} =50$, and
$N=5$.  The cluster search radii are $r=1.75$~\Mpc, and $r=2.75$~\Mpc,
for the parameter sets A and B, respectively. The parameter set A
finds the DF clusters in faint filaments crossing supercluster-defined
large voids, the parameter set B avoids most of the DF clusters in
these filaments.

The total numbers of the DF clusters found for the present epoch $z=0$ and 
the selection parameters used are given in Table~\ref{Tab1}. For the models
M256 and L100, we also give the numbers of the AHF halos.

\section{Analysis of models} 
 
\subsection{The density field of truncated models} 
 
To have an idea of the appearance of the density fields, we show in 
Fig.~\ref{fig:Mdenfield} the high-resolution density fields for all the
models of the series M256: M256.256, M256.064, M256.032, M256.016, 
M256.008, and M256.864.  To emphasise the filaments joining the DF clusters, 
we express densities in logarithmic scale, and use sheets of a 
thickness 6~\Mpc.  The fields are shown for the present epoch $z=0.0$. 
The models M768 and L256 look similar, when cuts at 
the same scales are used, but  fewer details are seen in the model M768
with respect to the model L256. 
 
Fig.~\ref{fig:Mdenfield} shows that the models M256.256 and M256.064 
display rather similar patterns of the cosmic web.  An even closer 
similarity is observed when we compare the density fields of the models 
L256.256 and L256.128.  This similarity of the patterns of the cosmic 
web in these models shows that the pattern of the web is  
defined by density perturbations of scales smaller than $\simeq 100$~\Mpc.

\begin{figure*} 
\centering 
\resizebox{0.25\textwidth}{!}{\includegraphics*{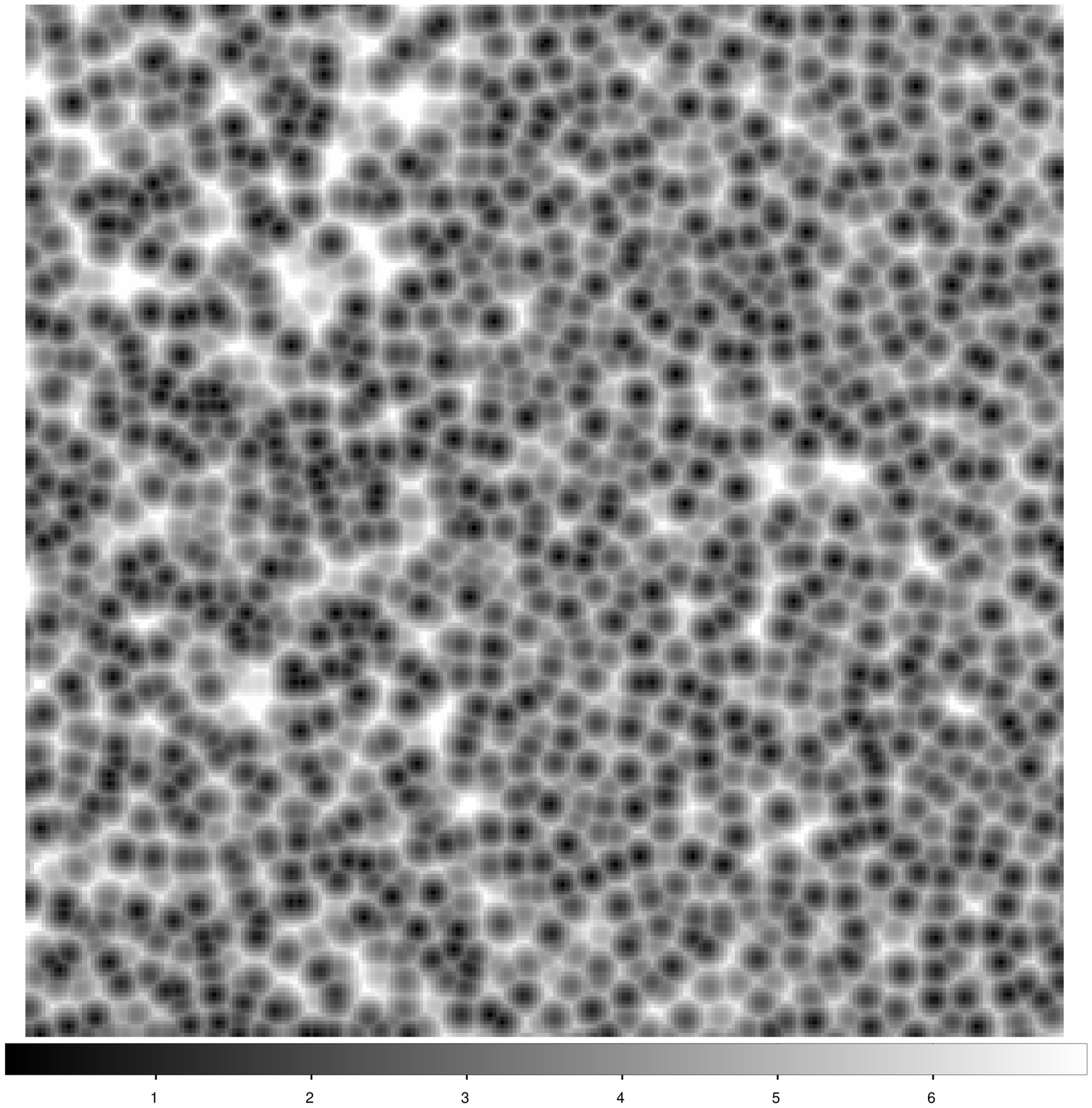}} 
\resizebox{0.25\textwidth}{!}{\includegraphics*{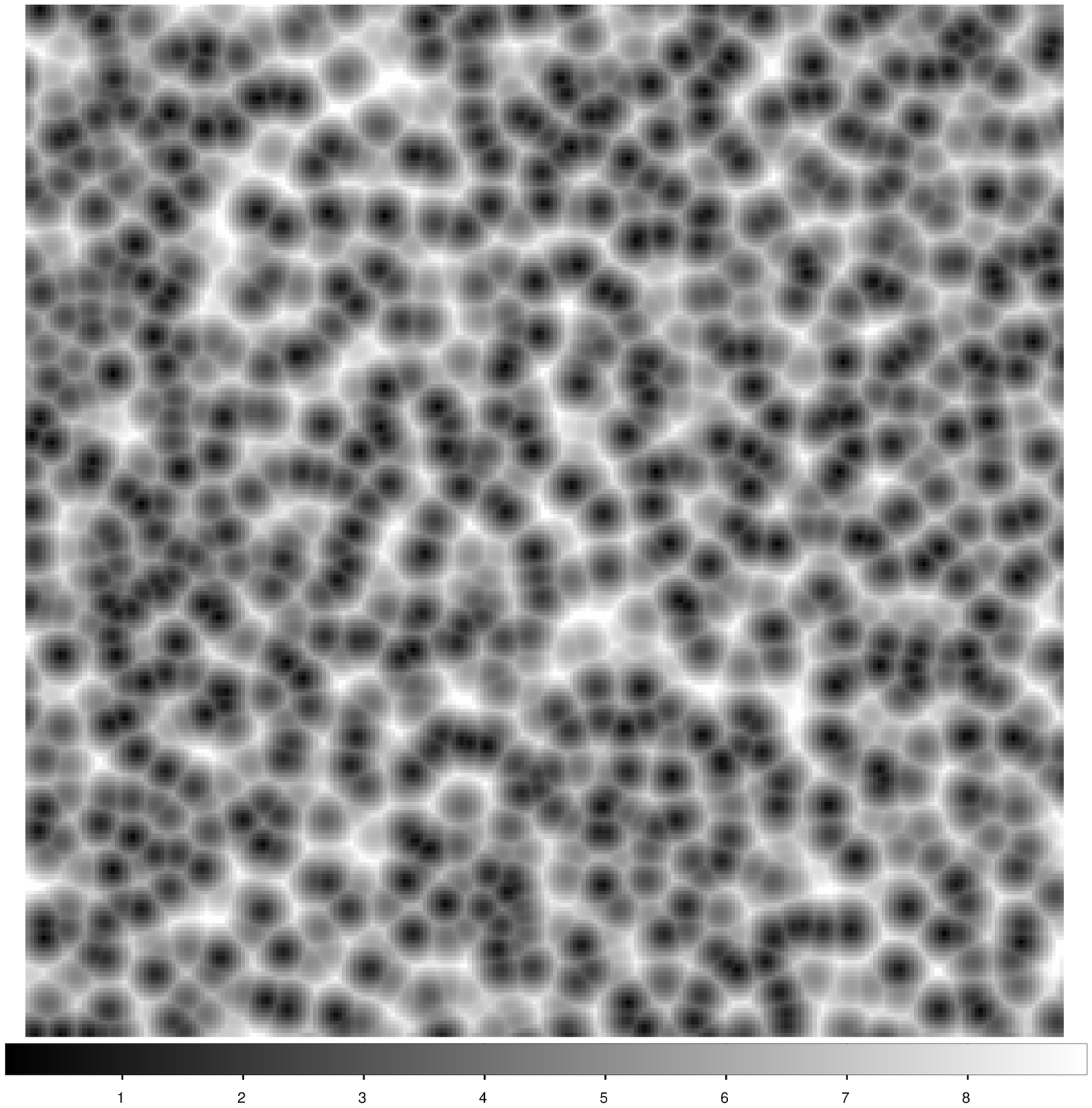}} 
\resizebox{0.25\textwidth}{!}{\includegraphics*{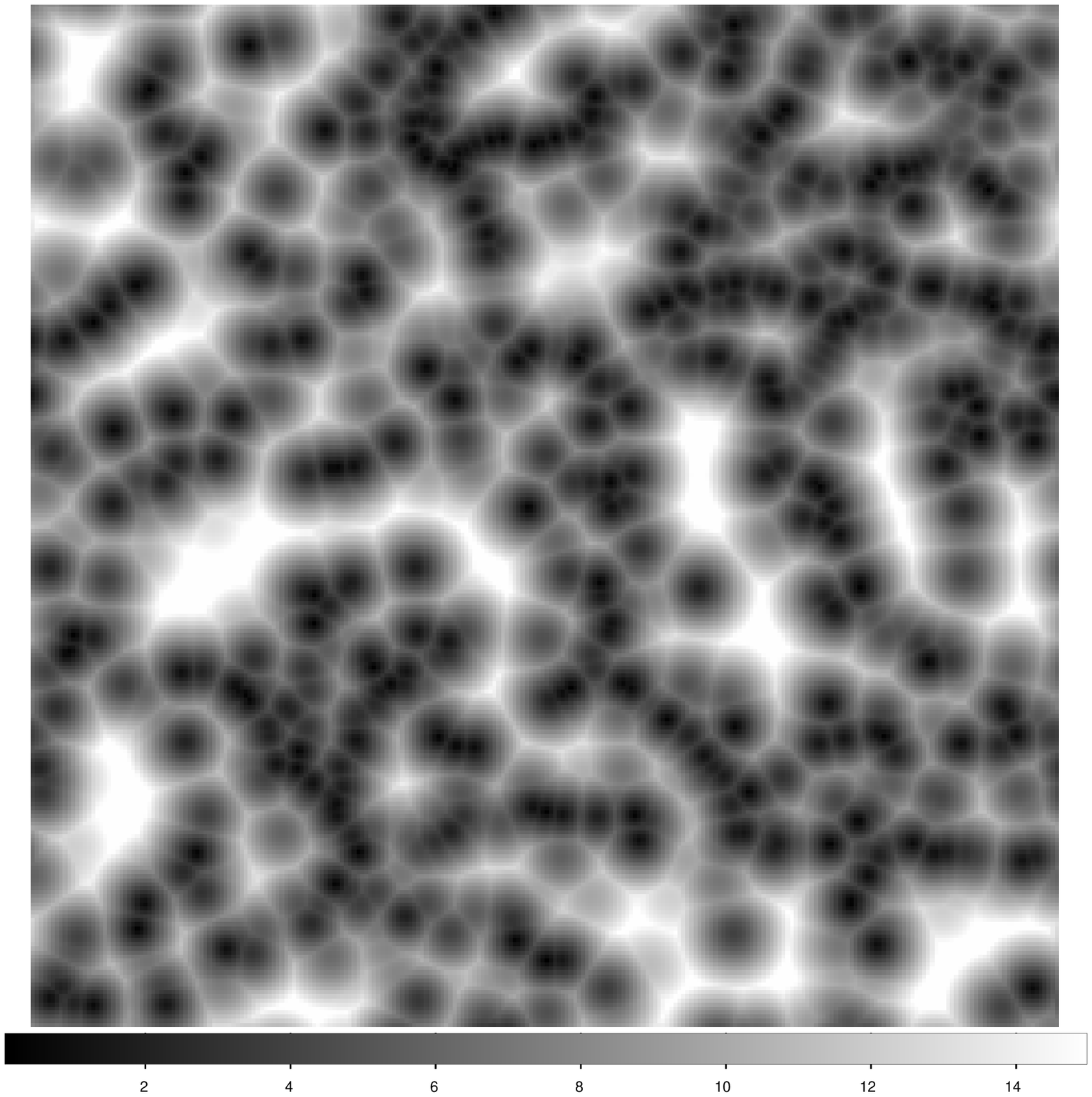}} 
\caption{The matrix of distances to the nearest DF clusters for the models
  M256.008, M256.016, and M256.256 (from left to right). The
  ``density'' is defined as the distance to the nearest DF
  cluster. The matrix is given for the $k=220$~\Mpc\ coordinate.  Note
  that changes of the mean radii in the distance matrix are similar to
  changes of the overdensity pattern of the density field, shown in
  Fig.~\ref{fig:Mdenfield}.  }
\label{fig:voidmatr} 
\end{figure*} 
 
In models with shorter scale power spectrum cuts large-scale features
disappear.  In the models with a cutoff scale $\leq 100$~\Mpc, the
shape of the pattern of the cosmic web is determined by the maximum
scale of density perturbations.  This is one of our main qualitative
conclusions: in the absence of large-scale perturbations, galaxy
systems larger than the cutoff scale do not form. In other words, the
scale of the cosmic web is determined by the density waves of the
largest scale present.  However, this conclusion is correct only up to
the scales $\sim 100$~\Mpc.  Fig.~\ref{fig:Mdenfield} (and similar
plots for other models) shows that the addition of perturbations
larger than this scale does not create larger systems, but only
amplifies them. We study this phenomenon below.

\subsection{The role of the density waves of medium scales} 
 
To see the effect of the absence of density perturbations of medium
scales, we performed one simulation, where the amplitude of the
initial density fluctuations between wavelengths 8~\Mpc\ and 64~\Mpc\
was truncated to zero.  This model (M256.864) is otherwise identical
to the model M256.256 and was generated using identical Fourier
amplitudes.  The power spectra of this model at $z=30$ and $z=0$ are
shown in Fig.~\ref{fig:spec}, and the density field at the present
epoch in Fig.~\ref{fig:Mdenfield}.

\begin{figure*}[ht]
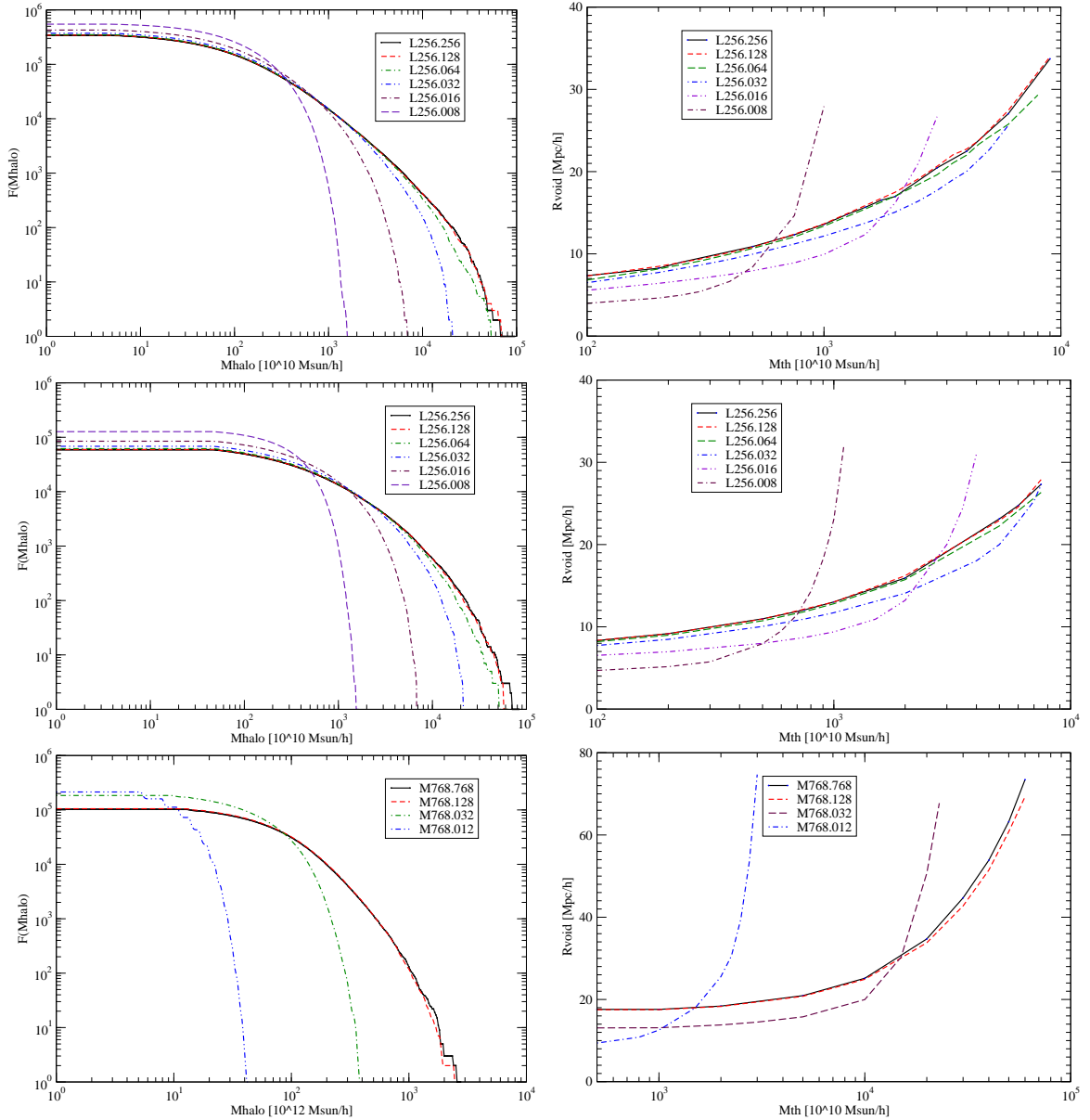
 
\centering 
\resizebox{0.40\textwidth}{!}{\includegraphics*{16394fg6a.eps}} 
\hspace{2mm}  
\resizebox{0.40\textwidth}{!}{\includegraphics*{16394fg6b.eps}}\\ 
\hspace{2mm} 
\resizebox{0.40\textwidth}{!}{\includegraphics*{16394fg6c.eps}} 
\hspace{2mm}  
\resizebox{0.40\textwidth}{!}{\includegraphics*{16394fg6d.eps}}\\ 
\hspace{2mm}  
\resizebox{0.40\textwidth}{!}{\includegraphics*{16394fg6e.eps}} 
\hspace{2mm} 
\resizebox{0.40\textwidth}{!}{\includegraphics*{16394fg6f.eps}}\\ 
\caption{The left panels show the cumulative mass functions of the
  density field clusters for the models with various cutoff
  scales. The right panels show the mean radii of voids, defined by
  the DF clusters for different threshold masses, $M_{\mathrm{th}}$,
  and for various cut scales. The upper panels are for the model L256
  with the DF cluster search parameter set A, the middle panels are
  for the same model with the DF cluster search parameter set B, the
  lower panels are for the model M768.  }
\label{fig:m256_clvoids} 
\end{figure*}

Fig.~\ref{fig:Mdenfield} demonstrates very clearly the role of density
perturbations of various scales in the formation of the cosmic web.
In the absence of large-scale perturbations, systems larger than the
cutoff scale do not form.  The result of removing only the
medium-scale perturbations is of particular interest.  As seen from
Fig.~\ref{fig:spec}, during the evolution the power spectrum on
intermediate scales, which is absent at the initial epoch, is in this
case almost restored.  However, this increase in the power spectrum
amplitudes does not lead to the formation of medium-scale galaxy
systems such as filaments.  In a model without the {\em initial}
medium-scale perturbations, filaments are absent, both within
superclusters and between them.  The distribution of small-scale
systems is more or less random, and there are no compact systems of
galaxies such as clusters -- the compact systems are rather small.
 
\subsection{The distribution of void sizes} 
 
Cosmic voids are regions of space devoid of certain kinds of objects
-- galaxies, clusters of galaxies etc.  Different types of objects
define voids of different size.  The reason for the dependence of void
sizes on the mass (or the luminosity) of objects used in their
definition is simple.  Large voids are determined by rich clusters and
crossed by filaments of faint galaxies.  Moreover, almost all systems
of galaxies contain outlying faint members (see Fig.~\ref{fig:dr7_240}
for the luminosity density field of a spherical shell of the Sloan
Digital Sky Survey).  Dwarf galaxies define much smaller voids than
giant ones \citep{Einasto:1986oh,Einasto:1989cr,Einasto:1991fq,
  Lindner:1995ui,Lindner:1996tu,
  Gottlober:2003,von-Benda-Beckmann:2008qf}. For the hierarchy of
voids, we also refer to \citet{van-de-Weygaert:1993},
\citet{Peebles:2001kl}, \citet{Gottlober:2003},
\citet{Aragon-Calvo:2007}, \citet{van-de-Weygaert:2009qf},
\citet{Aragon-Calvo:2010}, and \citet{Aragon-Calvo:2010wd}.

To find the distribution of void radii, we used a simple void finder
suggested by \citet{Einasto:1989cr}.  For comparison of different void
finders, we refer to \citet{Colberg:2008pd}. For each vertex of the
simulation grid, we first calculated its distance to the nearest DF
cluster.  For the positions of the DF clusters, we used the
$i,j,k$-indices of the maximum local density cells.  The distance
matrix is similar to the density field matrix; for the models
M256.008, M256.016, and M256.256, these matrices are shown in
Fig.~\ref{fig:voidmatr}.
 
Fig.~\ref{fig:voidmatr} shows the maxima of the distance matrix field.
These maxima correspond to the centres of voids, and their values are the
void radii.  The distribution of clusters is noisy, thus there are
many nearby local maxima in the distance matrix field.  We define the
position of the void centre as the location of the cell, which has the
highest distance value within a box of the size of $\pm 3$ grid
elements.
 
Fig.~\ref{fig:voidmatr} shows that in some places there are long ridges in 
the distance matrix; these ridges delineate elongated voids.  We 
consider them to be individual voids, if their centres are separated by more than 
3~\Mpc, i.e.\ they count as separate entries in the void search using the 
criterion shown above.  
 
In models with a higher cutoff, the maximum void radii increase with the 
cutoff scale, but only moderately.  There is only a small difference 
in the distribution of void radii between the models M256.256 and M256.064. In 
other words, very large density waves do not change the structure of 
the supercluster-void network, their role consisting essentially of the 
amplification of existing structures.

The mean void radii were found for a broad range of the DF cluster mass 
thresholds, $M_{\mathrm{th}}$.  The highest threshold 
used was selected so that the volume density of the DF clusters in the 
sample is approximately equal to the volume density of the Abell clusters, 
at about $25\times 10^{-6}$ (\Mpc)$^3$ \citep{Einasto:2006kl}. This gives 
$\simeq 420$ DF clusters in the volume of the L256 model.  The actual 
number of DF clusters used in the void definition depends on the value of 
the DF cluster mass threshold, $M_{\mathrm{th}}$, used in the void search. 
  
In Fig.~\ref{fig:m256_clvoids}, we show the mean void radii as
functions of the threshold mass $M_{\mathrm{th}}$ of the DF clusters
used in void search.  This figure shows that the void radii depend on
the threshold mass, as expected from similar studies of the void radii
in galaxy samples of different threshold luminosity, demonstrated in
many studies cited above.  We study this effect in more detail below.

\subsection{The distribution of the DF cluster masses} 
 
One of the characteristics of the cosmic web is the distribution of
cluster masses.  The cumulative mass functions of the DF clusters for
the models of the series L256 and M768 are shown in
Fig.~\ref{fig:m256_clvoids}.  We see that DF cluster masses strongly
depend on the scale of the power spectrum cutoff.  All models with a
cutoff on the scale 8~\Mpc\ have DF cluster mass distributions with
rather sharp decreases on the high mass side. The maximum masses of
the DF clusters in these models are $\simeq 1.5 \times
10^{13}~\mathrm{M}_\odot$ for the L256 model and $\simeq 4 \times
10^{13}~\mathrm{M}_\odot$ for the M768 model.  The maximum masses of
the DF clusters in the models of the L100 series are lower than the
maximum masses of the DF clusters in the models of the L256 series.
These differences are due to the larger sizes of the boxes used in the
DF cluster search in the models of the L256 and the M768 series.  As
shown in Fig.~\ref{fig:L100_massf}, there are practically no
differences between the maximum masses of the DF clusters and those
the AHF halos of the L100 model, since in the DF cluster search we
used a search radius approximately equal to the virial radius of the
most massive AHF halo.

With increasing spectrum cutoff scale, the mass distributions rapidly
shift to higher masses. This rapid increase in the maximum DF cluster
mass continues up to the cutoff scale of 64~\Mpc.  A higher cutoff
scale for the power spectrum only moderately increases the maximum
masses of the DF clusters.  This increase almost stops at the cutoff
scale 128~\Mpc.  The most massive DF clusters in these models have the
masses $\simeq 7 \times 10^{14}~\mathrm{M}_\odot$ and $\simeq 2 \times
10^{15}~\mathrm{M}_\odot$, for the models L256 and M768, respectively.
 
The number of low mass DF clusters in the models depends on the selection
parameters used in the cluster search.  The total number of DF
clusters in the model L256 found for the high-sensitive and the
low-sensitive parameter sets, A and B, is rather different. The
high-sensitive set A with a lower search peak density level
$D_{\mathrm{p}} =5$ has about six times more DF clusters than the set B
with $D_{\mathrm{p}} =50$.  All additional DF clusters found for the
parameter set A are in the low mass range, as seen in Table~\ref{Tab1}
and Fig.~\ref{fig:m256_clvoids}.  In the higher mass region, the cumulative
mass functions for both sets of search parameters almost coincide.  As
we see below, in the void search both parameter sets yield voids of
approximately equal mean radii for identical threshold masses.  This
shows that our results are rather robust to the parameter choice in
the DF cluster definition.

A more detailed study shows, however, that density waves of still
larger scales influence the structure of the cosmic web. We discuss
below the evolution of the density field in the void and supercluster
core regions.

\section{Discussion}

\subsection{Voids defined by clusters of different mass} 
 
As seen in Table~\ref{Tab1} and Fig.~\ref{fig:m256_clvoids}, the
numbers of the DF clusters for the search parameter set A for the models L256
are about six times larger than for the parameter set B.  The basic reason
for this difference is the use of a much lower threshold density $D_0=2$
and the peak density $D_{\mathrm{p}} =5$ in the cluster search; for the set B, the
corresponding densities are $D_0=10$, $D_{\mathrm{p}} =50$. For
instance, the number of the DF clusters in the L256.256 model for the
lowest mass threshold $10^{12}~\mathrm{M}_\odot$ is 149265 for the
parameter set A, and 59100 for set B. In spite of the large
difference in the number of the DF clusters for different search
parameters, the mean void radii for both parameter sets are
approximately equal, as seen in the right panels of
Fig.~\ref{fig:m256_clvoids}.  A detailed inspection of the density
field maps shows that this similarity in void sizes has a simple
reason: the DF clusters of the set B are high-density peaks inside the
same filaments that form the basic web of more densely populated
filaments of the DF clusters of the set A.  The density field maps also show
that there are no extra faint clusters inside the relatively small voids
defined by the clusters of the set A.  For void definition, it is
irrelevant how densely populated the filaments surrounding the voids are, if
there are no galaxies outside the filaments.
 
This conclusion agrees with the results of earlier studies by
\citet{Lindner:1996tu}, \citet{Peebles:2001kl}, and
\citet{Gottlober:2003} among others, that voids defined by
relatively faint galaxy filaments are completely devoid of galaxies.
 
Fig.~\ref{fig:m256_clvoids} shows that for a low DF cluster mass 
threshold, the void radii are almost independent of the mass threshold 
$M_{\mathrm{th}}$.  This means that in this DF cluster mass interval the
clusters are located in identical filaments. 
 
If the DF cluster mass threshold increases further, the void radii start
to increase.  This means that some filaments are fainter than the
respective mass threshold limit, and do not contribute to the void definition.
This slow increase in the mean void radii $R_{\mathrm{v}}$ as a function
of the DF cluster mass threshold $M_{\mathrm{th}}$ for strongly cut models
is valid until a certain $M_{\mathrm{th}}$ value, of about $3\times
10^{12}~\mathrm{M}_\odot$ and $10\times 10^{12}~\mathrm{M}_\odot$ for
the L256.008 and the L256.016 models, respectively.  Thereafter, with
an increasing mass threshold, the void radii increase very rapidly until
clusters disappear.  This effect is due to the very sharp decrease in
the number of DF clusters of a high mass (see the left panels of 
Fig.~\ref{fig:m256_clvoids}).  These rare clusters define very large
voids.  The sizes of these voids are not characteristic of the
overall cosmic web pattern of the particular model.
 
Fig.~\ref{fig:m256_clvoids} shows that in the models with a higher
cutoff scale $\lambda_{\mathrm{cut}} \geq 32$~\Mpc, the void mean
radii grow with the growth of the mass threshold smoothly, until the
highest masses, which correspond to very rich clusters.  Moreover, the
$R_v$ versus $M_{\mathrm{th}}$ curves for the models M256.032,
M256.064, M256.128, and M256.256 are almost identical.  This means
that the scale of the cosmic web is determined essentially by density
perturbations of a scale up to 32~\Mpc.  Some differences in the
$R_{\mathrm{v}}$ versus $M_{\mathrm{th}}$ curves remain between the
models M256.032 and M256.064.  The higher cutoff models are
practically identical in this void size test.
 
  Thus, the void analysis confirms our results from the mass 
  distribution of the DF clusters, that density perturbations of large 
  scales have little effect on the pattern of the cosmic web as 
  characterised by void sizes. 
 
The largest voids in the L256.256 model are defined by the rich clusters of 
mass threshold $9\times 10^{13}~\mathrm{M}_\odot$, and have the mean 
radii $\simeq 30$~\Mpc. These voids are characteristic of
filaments, which divide supercluster-defined voids (supervoids) into 
approximately equal sub-voids. 
 
The model M768 has a lower resolution; here faint filaments are
absent, and the scale of voids is defined by the DF clusters of higher
mass.  As in the model L256, models with a smaller cutoff scale have
smaller voids. The mean void radii for the models M768.128 and
M768.768 have practically identical $R_{\mathrm{v}}$ versus
$M_{\mathrm{th}}$ curves.  The largest voids in this model are defined
by the DF clusters of the mass threshold $6\times
10^{14}~\mathrm{M}_\odot$.  These very rich clusters populate the
cores of rich superclusters, thus the respective void radii are
practically equal to the radii of supervoids, $R_{\mathrm{v}} \simeq
70$~\Mpc\ \citep{Lindner:1995ui, Einasto:1997nx}.

\begin{figure*}[ht]
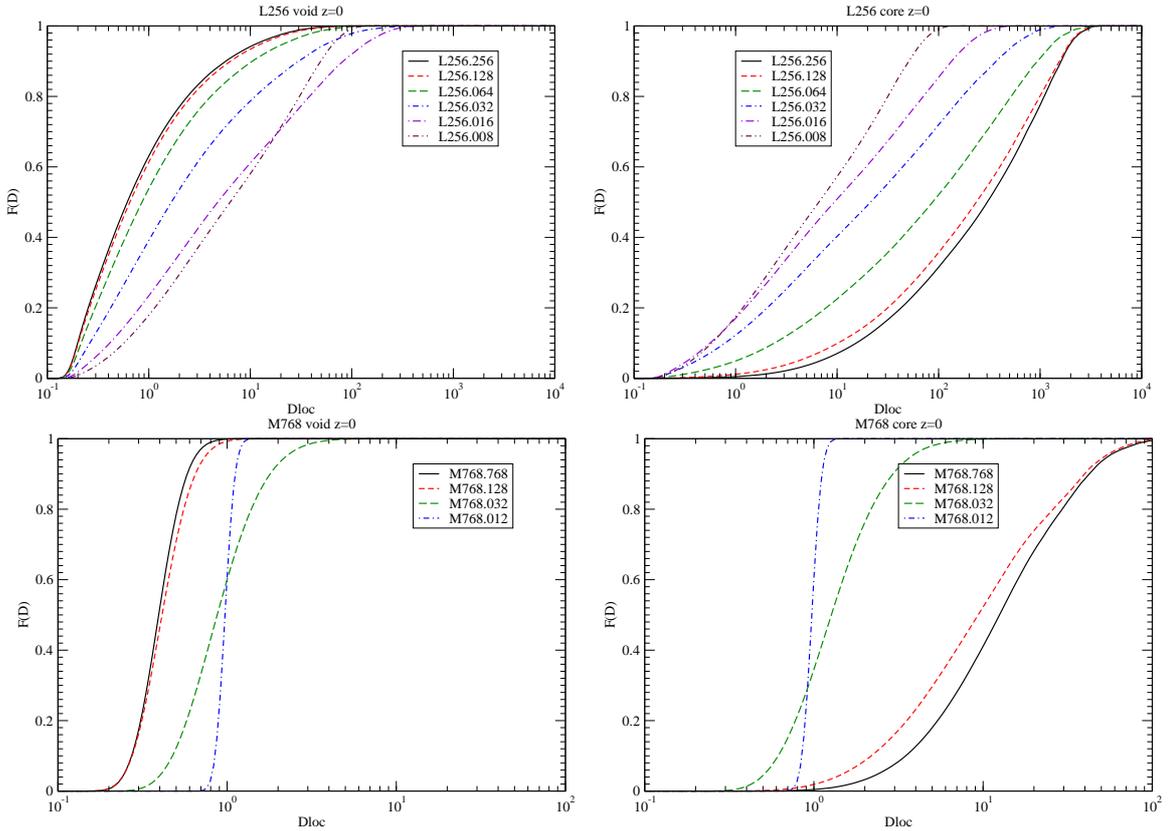
 
\centering 
\resizebox{0.40\textwidth}{!}{\includegraphics*{16394fg7a.eps}} 
\hspace{2mm}  
\resizebox{0.40\textwidth}{!}{\includegraphics*{16394fg7b.eps}}\\ 
\hspace{2mm}  
\resizebox{0.40\textwidth}{!}{\includegraphics*{16394fg7c.eps}} 
\hspace{2mm} 
\resizebox{0.40\textwidth}{!}{\includegraphics*{16394fg7d.eps}}\\ 
\caption{The cumulative distributions of the local densities for particles in
  the void and core regions for the present epoch $z=0$ are given in the
  left and the right panels, respectively. The upper row is for the
  L256 models, the lower row for the M768 models.  Models with various
  cutoff scale of the power spectra are shown.  }
\label{fig:dendistr} 
\end{figure*}

\subsection{The evolution of the density field in the void and 
  supercluster core regions}

As an additional test, we studied the evolution of the density field in 
extreme voids and supercluster cores.  Here we used the property of 
our models -- all variants of a series with a different spectrum cutoff 
scale were generated with an identical random number set.  Thus we can 
use the particle ID numbers to study the behaviour of the web on 
different spectrum cutoff scales. 
 
To define the extreme void and supercluster core regions we used the
global density field smoothed with a wider kernel, applying the same
procedure as explained above for finding the DF clusters in the
high-resolution density field.  To calculate the global density field,
we used for the models of the series L256 the kernel size 8~\Mpc, for
the models of the series M768 the kernel size 12~\Mpc\ when applying the
$B_3$ spline.  For every particle, we store the local density value at
the particle location (found on the basis of 27 nearest neighbours) in
addition to its coordinates, and the global density found from the
density field with a large smoothing length as described above. These
data are available for all models and evolution steps. We also found
the distributions of the local and global density of particles.

Using this information, we extracted particle samples for models with the full power
spectra at the present epoch, which had $\simeq 5$~\%
of particles with the lowest and the highest global density values. We
refer to these particle samples as the void and core samples, respectively.
The number of particles in the void and core samples of the M768.768 and
the L256.256 models are given in Table~\ref{Tab2}.  We calculated the
cumulative distributions of the local densities of the void and core particles at
the present epoch $z=0$ for all the models of the L256 and M768
series. The results are shown in Fig.~\ref{fig:dendistr}.
  
We discuss first the cumulative density distribution of the void and 
core particle samples of the M768 model.  This model has a lower 
resolution and lacks the low-mass DF clusters in the deep void regions, so the 
interpretation of the results is simpler.  Fig.~\ref{fig:dendistr} 
shows that the cumulative particle density distributions of the void and core 
regions of the model M768.012 are almost identical. Both distributions 
are rather symmetrical around the mean density level $D_{\mathrm{loc}} = 
1$.  This result shows that in this model, there is no difference 
between the void and core regions. The symmetry around the mean density 
level indicates that the evolution is still close to the linear regime 
of the growth of density perturbations, i.e the growth of the negative and 
positive sections of the density contrast $\delta = D - 1$ is similar 
and proportional to the growth factor of the evolution. 
 
In the model M768.032, the amplification of the density growth by the density 
waves of medium scale is already visible. The difference between the
density distributions of the void and core samples is still small -- there 
are no big voids and no rich supercluster cores.

The distributions of the particle densities in the void and core regions are 
completely different in the models M768.128 and M768.768.  In the void regions, 
all local densities are lower than the mean density, i.e.\  there 
are no systems of galaxies. We recall that for the formation of a 
galaxy or a galaxy group the local density must exceed a certain threshold, 
about 1.6 in the mean density units.  In contrast, in the core regions there 
are almost no particles of local densities less than the mean 
density. 
 
\begin{table}[ht] 
\caption{Results of the K-S test.} 
\begin{tabular}{lrrr}  
\hline  
Model   & $n$ & $d$    & $p$   \\  
\hline  
\\ 
M768 void    &  820002 &  0.07563 & 0.000000   \\ 
M768 core    &  823717 &  0.12304 &  0.000000  \\ 
\\ 
L256 void    & 6677929 &  0.01747 & 0.000000      \\ 
L256 core    & 6768219 &  0.04725 &  0.000000     \\ 
\\ 
\label{Tab2}                         
\end{tabular} 

The meaning of $n$, $d$ and $p$ is explained in text. 
\end{table} 
 
We also note that there is a small but definitely significant 
difference between the distribution of densities in the void regions of the
models M768.128 and M768.768, and in the core regions of the same models. 
The difference is smaller in the void regions, but in both regions it is 
clearly present. The results of the Kolmogorov-Smirnoff test of the 
comparison of the cumulative distributions of particle densities in the void (core) 
 regions of the M768.128 versus M768.768 and the L256.256 versus 
L256.128 models are shown in Table~\ref{Tab2}, where $n$ is the number 
of particles in respective samples (which are identical in the full model and 
the 128~\Mpc\ cut models of the same series), $d$ is the maximum 
difference of cumulative distributions, and $p$ is the probability 
that the distributions compared are taken from the same parent sample. As 
we see, this test shows that the distributions are different at high 
significance levels.  This shows that the density perturbations of a
larger scale than 
128~\Mpc\ make voids emptier and systems of galaxies richer, i.e.\ they 
amplify the emptiness and the richness of the cosmic web.

The model L256 has much higher spatial and mass resolutions.  For 
this reason, there are more differences between the particle local 
density distributions of the void and core regions.  In this model in the void 
regions we see the presence of particle overdensities,  with local 
densities $D > 1$.  These particles form faint filaments crossing 
large voids. It is important that the larger the power spectrum 
cutoff scale (i.e.\ the larger perturbations are included in the 
simulation), the lower the fraction of the number of particles with 
overdensities.  In other words, large-scale perturbations suppress the 
formation of filaments in void regions. 
 
In the core regions the growth of overdensities is more rapid, the
larger is the scale of the perturbations included in simulations. In
the model L256.256, there are no particles with local densities below
the mean density, i.e.\ all particles have been moved to rich
systems. Owing to the higher resolution, the highest densities in the
core regions are much higher than in the model M768.

\subsection{Why the perturbations of the largest scales do not increase the
  scale of the cosmic web?}
 
Our study suggests that density perturbations up to the scale $\simeq 
100$~\Mpc\ determine the scale of the cosmic web in terms of void 
sizes.  In contrast, waves of larger wavelengths do not influence the scale of 
but only amplify the web, leaving its scale unaffected.  Thus, it is unclear, 
why the growth of the scale of the skeleton of the 
cosmic web with increasing wavelengths of density perturbations, stops at the
perturbation scale $\simeq 100$~\Mpc.  Has this some deeper physical 
meaning? That the change in the behaviour of the waves of 
different scale occurs in numerical simulations where 
only standard physics of the early Universe is included, implies that there is a 
simple answer.
 
For $h\approx 0.7$, this largest pattern scale is close to
$R_{\mathrm{eq}}\equiv a(t_0)\eta_{\mathrm{eq}}=2(\sqrt{2} -1)
(c/H_0)\times \sqrt{\Omega_{\mathrm{rad}}}/\Omega_{\mathrm{m}} \approx
120$ Mpc, which is the only scale characterising a primordial
scale-free ($n_{\mathrm{s}}\approx 1$) spectrum of density
perturbations multiplied by the scale-dependent transfer function
arising due to the transition from the radiation-dominated stage to
the matter-dominated one, which occurred at $z=z_{\mathrm{eq}}\approx
3200$ according to the most recent observational data (see
\citet{Jarosik:2010}). Here the conformal time
$\eta_{\mathrm{eq}}=\eta(z_{\mathrm{eq}})=c\int_{z_{\mathrm{eq}}}^\infty
\, dz/H(z)$, $H\equiv \dot a(t)/a(t)$ is the Hubble parameter, $t_0$
is the present time ($H(t_0)=H_0$), and $\Omega_{\mathrm{rad}}$
includes the contribution from three species of neutrino which all may
be considered massless at the moment of matter-radiation
energy-density equality $z=z_{\mathrm{eq}}$. We note that
$R_{\mathrm{eq}}$ does not depend on $h$ if expressed in terms of
$\Omega_{\mathrm{rad}}h^2\propto T_{\gamma}^4$ and
$\Omega_{\mathrm{m}}h^2$, nor depend on the dark energy equation of
state $w_{\mathrm{DE}}$ ($w_{\mathrm{DE}}=-1$ for a cosmological
constant), only the present value of $\Omega_{\mathrm{m}}$ being
important.  Moreover, the expression for $R_{\mathrm{eq}}$ remains the
same for an open (negatively spatially curved) Universe in the absence
of a cosmological constant.
 
Additional insight to this problem is given by simulations designed to
understand the possible future of the development of the cosmic web.
These simulations were performed by a number of authors
\citep{Loeb:2002sh,Nagamine:2003sj, Busha:2005le, Dunner:2006,
  Dunner:2007, Hoffman:2007dd,Krauss:2007wb}.  In the early Universe,
it is well known that the matter density $\Omega_{\mathrm{m}}$
dominated over the energy density. As time proceeded and the Universe
expanded, the $\Omega_\Lambda$ term increased and caused the Universe
to expand in an accelerating fashion from a redshift of $z \simeq
0.8$.

Our results for the absence of the growth of the scale of the cosmic
web with the increase in the scale of density perturbations over
$\simeq 100$~\Mpc\ can probably be explained as the result of the
freezing of the web that started at recent redshifts, $1+z <
((1-\Omega_{\mathrm{m}})/\Omega_{\mathrm{m}})^{1/3}$ for an exact
cosmological constant.  Simulations of the future development of the
web mentioned above support this interpretation. As stated by
\citet{Hoffman:2007dd} ``in comoving coordinates the future
large-scale structure will look like a sharpened image of the present
structure: the skeleton of the cosmic web will remain the same, but
clusters will be more isolated and the filaments will become
thinner''.  Simulations by \citet{Nagamine:2003sj} and
\citet{Hoffman:2007dd} among others have shown that the mass evolution
of bound objects such as clusters will stop. In the very distant
future, all galaxies that are not bound to the Local Group (which
consists basically of the merged Milky Way and M31) will fade away
while approaching the event horizon, as seen from an observer inside
the Local Group. It remains to be investigated how a possible
deviation of $w_{\mathrm{DE}}$ from $-1$, if it exists, can affect the
freezing and the properties of the cosmic web.

\section{Conclusions} 
 
The basic conclusions of our study are as follows: 
 
\begin{itemize} 
\item{} The properties of the cosmic web depend strongly on density 
  perturbations of various scales. 
 
\item{}  Small-scale perturbations up to the scale $\simeq 8$~\Mpc\ are 
  responsible for the formation of galaxy and cluster type systems. 
 
\item{} Medium-scale perturbations of the scale $8 \dots 64$~\Mpc\ form 
  the filamentary web inside and between the superclusters. They also 
  contribute to the amplification of the systems formed by perturbations 
  of smaller scales. 
 
\item{}  The cosmic web with filamentary superclusters and voids is 
  formed by the combined action of all perturbations up to the scale $\simeq 
  100$~\Mpc.  The largest perturbations in this range determine the 
  scale of the supercluster-void network. 
 
\item{} Perturbations of the largest scales $>100$~\Mpc\ modulate the 
  richness of galaxy systems from clusters to superclusters, and make 
  voids emptier.

\end{itemize} 
 
\begin{acknowledgements} 
 
  We thank the anonymous referee for stimulating suggestions.  The
  present study was supported by the Estonian Science Foundation
  grants No.  7146 and 8005, and by the Estonian Ministry for
  Education and Science grant SF0060067s08. The study has also been
  supported by ICRAnet through a professorship for Jaan Einasto, and
  by the University of Valencia (Vicerrectorado de Investigaci\'on)
  through a visiting professorship for Enn Saar and by the Spanish MEC
  projects ``ALHAMBRA'' (AYA2006-14056) and ``PAU'' (CSD2007-00060),
  including FEDER contributions. J.E., I.S. and E.T.  thank
  Astrophysikalisches Institut Potsdam (using DFG-grant Mu 1020/15-1),
  where part of this study was performed.  J.E. thanks also the Aspen
  Center for Physics and the Johns Hopkins University for hospitality
  where this project was started and continued.  The simulation for
  the model L256 was calculated at the High Performance Computing
  Centre, University of Tartu. In plotting of density fields we used
  the SAOImage DS9 program.  A.A.S. acknowledges the RESCEU
  hospitality as a visiting professor. He was also partially supported
  by the Russian Foundation for Basic Research grant No. 11-02-00643
  and by the Scientific Programme ``Astronomy'' of the Russian Academy
  of Sciences.

  We thank the SDSS Team for the publicly available data releases. 
  Funding for the SDSS and SDSS-II has been provided by the Alfred 
  P. Sloan Foundation, the Participating Institutions, the National 
  Science Foundation, the U.S. Department of Energy, the National 
  Aeronautics and Space Administration, the Japanese Monbukagakusho, 
  the Max Planck Society, and the Higher Education Funding Council for 
  England. The SDSS Web Site is \texttt{http://www.sdss.org/}.

The SDSS is managed by the Astrophysical Research Consortium for the 
Participating Institutions. The Participating Institutions are the 
American Museum of Natural History, Astrophysical Institute Potsdam, 
University of Basel, University of Cambridge, Case Western Reserve 
University, University of Chicago, Drexel University, Fermilab, the 
Institute for Advanced Study, the Japan Participation Group, Johns 
Hopkins University, the Joint Institute for Nuclear Astrophysics, the 
Kavli Institute for Particle Astrophysics and Cosmology, the Korean 
Scientist Group, the Chinese Academy of Sciences (LAMOST), Los Alamos 
National Laboratory, the Max-Planck-Institute for Astronomy (MPIA), 
the Max-Planck-Institute for Astrophysics (MPA), New Mexico State 
University, Ohio State University, University of Pittsburgh, 
University of Portsmouth, Princeton University, the United States 
Naval Observatory, and the University of Washington.

\end{acknowledgements}


\end{document}